\documentclass[%
 reprint,
amsmath,
amssymb,
aps,
 floatfix,
]{revtex4-2}

\usepackage{graphicx}
\usepackage{dcolumn}
\usepackage{bm}
\usepackage{qcircuit}
\usepackage{physics}
\usepackage{xcolor}
\usepackage[colorlinks]{hyperref}
\usepackage[capitalize]{cleveref}
\usepackage{mathtools}
\usepackage{tabularx}
\usepackage{multirow}
\usepackage[caption=false]{subfig}
\usepackage{amsmath}
\usepackage{algpseudocode}
\usepackage{algorithm}

\begin{document}

\title{Resource-efficient context-aware dynamical decoupling embedding for arbitrary large-scale quantum algorithms}

\author{Paul Coote}
\author{Roman Dimov} 
\author{Smarak Maity}
\author{Gavin S. Hartnett}
\author{Michael J. Biercuk}
\author{Yuval Baum}
\email{yuval.baum@q-ctrl.com}
\affiliation{Q-CTRL, Los Angeles, CA USA and Sydney, NSW Australia}

\date{\today}

\begin{abstract}
We introduce and implement GraphDD: an efficient method for real-time, circuit-specific, optimal embedding of dynamical decoupling (DD) into executable quantum algorithms. We demonstrate that for an arbitrary quantum circuit, GraphDD exactly refocuses both quasi-static single-qubit dephasing and crosstalk idling errors over the entire circuit, while using a minimal number of additional single-qubit gates embedded into idle periods. The method relies on a graph representation of the embedding problem, where the optimal decoupling sequence can be efficiently calculated using an algebraic computation that scales linearly with the number of idles. This allows optimal DD to be embedded during circuit compilation, without any calibration overhead, additional circuit execution, or numerical optimization. The method is generic and applicable to any arbitrary circuit; in compiler runtime the specific pulse-sequence solutions are tailored to the individual circuit, and consider a range of contextual information on circuit structure and device connectivity. We verify the ability of GraphDD to deliver enhanced circuit-level error suppression on 127-qubit IBM devices, showing that the optimal {\em circuit-specific} DD embedding resulting from GraphDD provides orders of magnitude improvements to measured circuit fidelities compared with standard embedding approaches available in Qiskit.
\end{abstract}

\maketitle

\section{Introduction}

Dynamical decoupling (DD) is an open-loop quantum control method that suppresses errors by repeatedly reversing the sense of error accumulation \cite{Vitali1999, Viola1998, DUAN1999}. Periods of forward error accumulation can be offset against equal periods of reversed error accumulation, in the manner of a spin-echo \cite{hahn1950spin}. This process is known as `refocusing' of errors, and typically involves the application of a timed sequence of single-qubit unitary operations during periods of idle time evolution. Crucially, the utility of decoupling does not depend on precise knowledge of the rates of error accumulation; it requires only a reliable way to reverse these rates at chosen moments \cite{Viola1999}. This technology has been explored extensively in the context of quantum computing, and demonstrated to dramatically extend the lifetime and fidelity of idling qubits~\cite{Jurcevic_2021,Pokharel2018, Das2021,ezzell2023dynamical,Uhrig2007,Quiroz2013,Biercuk2009}.

Going beyond individual isolated qubits, DD has been identified and validated as a method for improving circuit fidelities in quantum computing~\cite{lubinski2024,Mundada_2023,kam2023,pelofske2024,Pokharel_2023}.  In such settings DD can do more than suppress single-qubit dephasing: harkening back to its original use in NMR \cite{shaka1987broadband,coote2018}, DD can also be used to suppress crosstalk errors induced by quasistatic couplings~\cite{Paz-Silva_2016,niu2024} or couplings induced by activation of gates on proximal qubits. Several approaches exist for suppressing crosstalk errors and phase errors within relatively simple contexts, such as when two adjacent qubits are mutually idle for exactly the same interval \cite{tripathi2022suppression, niu2022analyzing,niu2024,evert2024}. These schemes are of limited use for a generic input circuit.

The efficacy of DD in a generic circuit depends jointly on the decoupling sequences applied to multiple qubits, and on the context in which the sequences are applied~\cite{niu2022analyzing,Seif_2024}. High-performance decoupling protocols, useful in generic quantum circuits, must respect the complex structure of arbitrary quantum algorithms and their compiled implementations. For instance, in a typical quantum circuit, delay commands are distributed by the compiler throughout the circuit; the various qubits become idle or active at different times, and periods of mutual idling (when crosstalk errors arise) may begin and end asynchronously amongst the many pairs of coupled qubits. A particular qubit may in general be subjected to crosstalk from multiple neighbors at the same time, and the error terms accumulate at different unknown rates and over various temporal intervals that may be fully or partially aligned.  Given these complications, the process of optimal DD embedding is highly dependent on the circuit (algorithm) and device. In particular, the ideal decoupling sequence cannot be derived once and then applied equivalently to different circuits. To the best of our knowledge, there is no previous work that solves the optimal embedding problem for a generic circuit. Instead, existing automated DD embedding schemes imperfectly suppress crosstalk, and/or rely on empirical tuning that has high overhead and is not scalable \cite{pokharel2023Algorithmic,Das2021,Xie2022,pelofske2023scaling,saki2023hypothesis,Jurcevic_2021,Ji_2024,Niu_2022effects,tong2024}.

In this work we present GraphDD \cite{DD_patent}: a solution to the {\em circuit-specific} DD embedding problem. GraphDD is optimal, in that it finds a circuit-wide decoupling protocol that exactly suppresses all phase and crosstalk errors for a completely arbitrary configuration of idle delays in any quantum circuit.  GraphDD is automatic, efficient, and economical with the number of embedded gates (using the minimum of two gates per idle where it is possible to do so). This is achieved by finding a favorable {\em ordering} of the idles, and embedding gates in each idle according to that ordering. Our novel representation of the embedding problem allows the ordering to be determined using the structure and properties of a computational graph which represents the DD embedding task, and is constructed from the specific details of the input circuit and hardware device. Given this ordering of idles, optimal embeddings (with complete suppression of phase and crosstalk errors) can be found for arbitrary input circuits, using a calculation that scales linearly with the number of idle delays in the circuit. We compare this method to the native Qiskit DD embedding protocol and execute on IBM quantum computers. Experimental tests on identical circuits (except for the DD embedding scheme) show up to $200\times$ and over $9,000\times$ improvement in the success probability of the Quantum Fourier Transform and Bernstein-Vazirani algorithms, respectively.  We show that using a metric of selectivity, measuring the ``signal-to-noise'' of the deterministic circuit output, the effective useful circuit width can be increased over $3\times$ ($4\times$) for Quantum Fourier Transform (Bernstein Vazirani).

\section{Theory and implementation}

\subsection{Contextual decoupling}

Idling error during quantum circuits can introduce a substantial degradation of circuit-level performance that is not accounted for by gate-level fidelity metrics~\cite{Mundada_2023}. The major sources of idling error are single qubit dephasing (with characteristic timescale $T_2$) and parasitic crosstalk amongst pairs of qubits. Such errors can be largely suppressed using {\em dynamical decoupling}---the strategic insertion of additional gates into the idles, arranged in such a way as to suppress the dominant coherent error sources for idle qubits. 

In order to exploit this well-tested physical approach in the context of complete quantum circuits, it is important to distinguish two different problems: (i) context-{\em specific} DD versus (ii) DD embedding for {\em any arbitrary} circuit. Designing a decoupling scheme for a {\em particular} context, such as preserving a single-qubit state for as long as possible, has received a great deal of attention in the research literature \cite{Khodjasteh2013,Pokharel2018, Das2021,ezzell2023dynamical,Uhrig2007,Quiroz2013, Biercuk2009}. Similarly, many appropriate multiqubit DD motifs are known for specific contexts, such as two otherwise isolated qubits that are idle for the same duration~\cite{Paz-Silva_2016}.

A compiled quantum circuit, in general, contains many idle delays which possess different durations and contexts with regard to actions applied to neighboring qubits in the circuit.  Given a compiled quantum circuit with idle periods, and an error Hamiltonian that describes the accumulation of coherent errors during the idles, the generalized DD embedding problem is to determine the placement of additional gates to most effectively suppress the error Hamiltonian while preserving the logical operation of the circuit.  

A conventional error Hamiltonian contains phase accumulation ($Z$) and cross talk ($ZZ$) errors, as encountered in e.g. superconducting circuits. The coefficients of these error terms are considered to be quasi-static: time-invariant at least on a time-scale comparable to the idles' durations. This approximation is sufficient for appropriate separation of timescales typically encountered in real quantum hardware subject to drift dynamics~\cite{Carvalho2021}. 

More precisely, the error Hamiltonian is
\begin{align}
    H_{\text{error}} = \sum_k \epsilon_k(t) Z_k + \sum_{\langle j,k \rangle}J_{j,k} Z_jZ_k \label{eq:hamiltonian}
\end{align}
where $\epsilon$'s and $J$'s are unknown (or imprecisely known). The second sum is over qubits $j$ and $k$ that are connected on the device; that is, crosstalk interactions arise between coupled qubits. Although the $\epsilon$'s are time-dependent, they are approximately constant on the time-scale of a typical idle period; $\epsilon_k(t)$ is a stochastic function possessing temporal correlation (i.e., it is a non-Markovian process) with autocorrelation exhibiting a characteristic timescale $T_2$. That noise correlation is what allows the use of DD to suppress phase errors: DD is effective, provided that the frequency of decoupling is large compared to $1/T_2$. In this regime, we can consider the parameters of Eq.~\eqref{eq:hamiltonian} to be quasi-static. Single-qubit phase accumulates at rate $\epsilon_k$ when qubit $k$ is idle. Similarly, ZZ-phase (crosstalk) accumulates at rate $J_{j,k}$ only when qubits $j$ and $k$ are mutually idle over the same temporal interval. Although these error sources can also be present during gates (i.e. when qubits are {\em not} idle), we assume that a separate gate optimization procedure addresses all errors within the context of a gate (for example, the echo structure of the ECR gate is designed to be robust to crosstalk errors within the gate duration \cite{Sundaresan_2020}, or autonomously designed gates incorporate such dynamics natively~\cite{Baum2021}).

\begin{figure}
\centering
\includegraphics[width=0.95\linewidth]{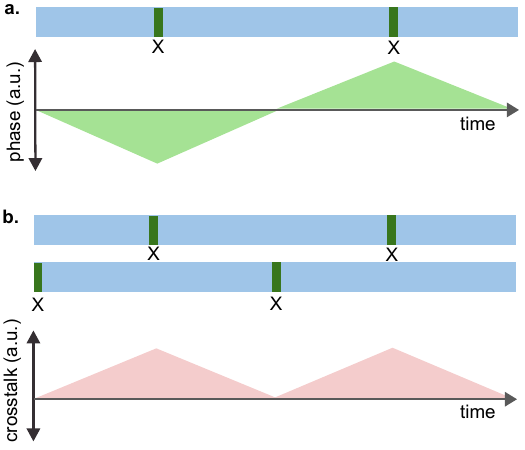}
\caption{\label{fig:errors} Refocusing of idling errors. (a) During an idle delay (blue), phase error grows at an unknown rate (green polygon), but is refocused using two X-gates. (b) Crosstalk error (red polygon) also grows at an unknown rate, and is refocused by offsetting the DD sequences on the two relevant qubits. }
\end{figure}

Fig.~\ref{fig:errors} illustrates the underlying physical principle behind error refocusing in the context of DD embedding. During an idle delay, phase error grows at an unknown rate. The sense of phase accumulation can be reversed by an X-gate (or any $\pi$-pulse). This is a direct consequence of the anti-commutation of the Pauli matrices, for instance $XZX = -Z$. Therefore, two X-gates, separated by 50\% of the idle duration, will leave the phase error at exactly zero by the end of the idle period. Selecting two X-gates ensures that the error suppressing sequence results in a net logical identity operation (Fig.~\ref{fig:errors}a). Beyond single-qubit dynamics, in the case of mutually idle qubits, crosstalk error can also be refocused using X-gates applied to both qubits. In this case, gates applied simultaneously on both qubits do {\em not} reverse the crosstalk accumulation, since $X_1X_2\,Z_1Z_2\,X_1X_2 = +Z_1Z_2$. However, an X-gate applied on just {\em one} qubit can reverse the sense of the error; i.e., $I_1X_2\,Z_1Z_2\,I_1X_2 = X_1I_2\,Z_1Z_2\,X_1I_2 = -Z_1Z_2$. Therefore, the two sets of two X-gates on the two qubits must be offset temporally and applied with appropriate (anti)symmetrization of the system evolution~\cite{Paz-Silva_2016, evert2024, niu2024}. This approach will return both the phase errors and the crosstalk error to zero at the end of the mutually idle period (Fig.~\ref{fig:errors}b). For simplicity of analysis, we treat X-gates as instantaneous and ideal, but various mechanisms exist to consider the impact of nonzero pulse-duration on sequence performance~\cite{Green_2013}. 

In the remainder of this section, we build on these well established physical concepts describing controlled evolution in average-Hamiltonian theory in order to produce an efficient embedding technique for DD protocols into arbitrary quantum circuits.  We present key elements in a logical order motivated by physical understanding and follow with presentation of a summary algorithm which explains the full embedding procedure.

\begin{figure*}
\centering
\includegraphics[width=0.95\linewidth]{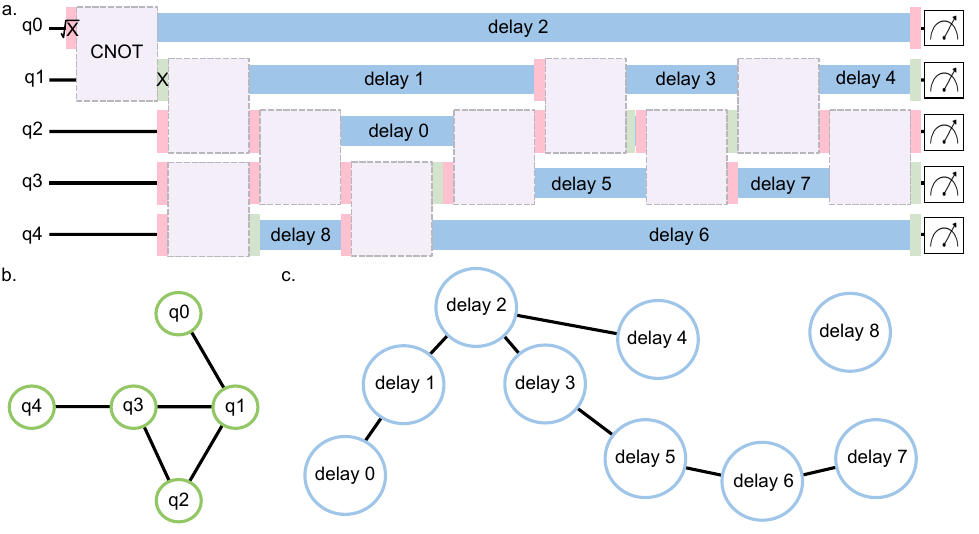}
\caption{\label{fig:dd_embedding_graph} DD embedding graph representation. (a) Schematic of a scheduled circuit. Single and two qubit gates are represented by narrow and wide boxes, respectively. Idle delays are distributed asynchronously throughout the circuit by the compiler and are numbered from 0 to 8. (b) A device connectivity graph indicates the couplings between qubits. It is a property of the device and does {\em not} change for different input circuits. (c) A graph representing the DD embedding problem. This graph depends on the specific input circuit. In the DD graph, the number of nodes is equal to the total number of delay instructions in the circuit schedule (it is not directly related to the number of qubits in the algorithm or the full device). The number of edges is partly dependent on the device connectivity, but also on the temporal overlap of the various idle delays. In this example, the graph is acyclic, and it comprises two connected subgraphs. 
}
\end{figure*}

\subsection{Graph representation of DD embedding}
\label{subsec:DD_as_a_graph}
The first step in developing a methodology for generalized DD embedding leveraging these insights is to find a useful representation of the structure of the problem.  A pathway forward emerges by examining the specific physical operations and objectives that must be undertaken.  The relevant information for determining the embedding within a particular idle delay includes: its own timing information (start and stop times); the timing information of any other idles that overlap in time and are physically coupled to the current idle's qubit; and, the DD gates (and their timing) of any other idles that overlap in time and are coupled to the currently idle qubit.

These characteristics and their interdependencies can be represented by a computational graph. Each idle delay in the circuit is represented by a node, and each edge represents potential crosstalk channels, connecting two nodes that overlap in \emph{both time and device connectivity}. Fig.~\ref{fig:dd_embedding_graph} depicts a scheduled quantum circuit and the graph that represents the delays and potential crosstalk errors. It is important to note that this graph is distinct from the device connectivity graph, and from the directed acyclic graphs that are sometimes used to represent quantum circuits. 

\begin{figure} [t!]
\centering
\includegraphics[width=0.95\linewidth]{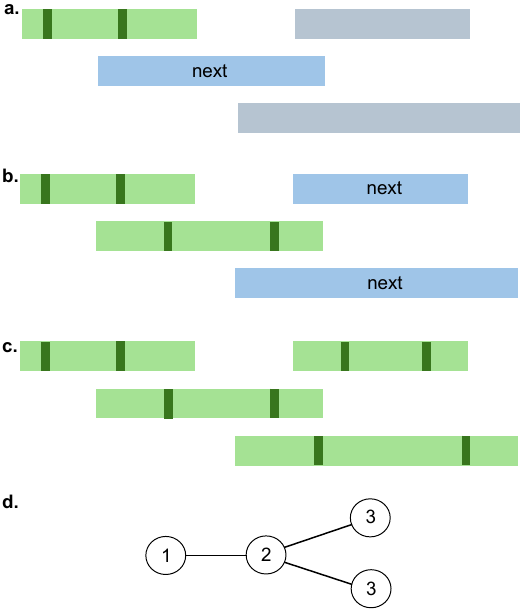}
\caption{\label{fig:gateset_acyclic} An illustration of the graph traversal. (a-c) show four idling that that may appear in a circuit (gates are not shown). Green idles are visited nodes, i.e., decoupling gates were already embedded, blue idling represent the current node for which we wish to find embedding. In (a) one idle has its DD gates already chosen (green), and another three idles are yet to be addressed. In this example, crosstalk is possible between adjacent qubits. (b-c) By addressing the idles in a particular order, we can ensure that each idle's embedding depends on at most one other idle's previously chosen gate positions. In this case, both blue periods can be addressed simultaneously or in any order. (d) The embedding ordering is an either depth-first or breath-first traversal of the DD embedding graph. Since the graph is acyclic, during the traversal each current node has at most one immediate ancestor.
}
\end{figure}

Once we have the problem graph, our overall embedding strategy is to traverse this graph, embedding X-gates in each node (idle) in turn. A graph {\em traversal} is a procedure for visiting all nodes of a given graph in a specific order. The ordering is determined from the structure of the graph by a specific traversal algorithm.  Nodes visited {\em before} the current node are called {\em ancestors}. Ancestors that are connected to the current node are called {\em direct} ancestors. Acyclic graphs can be traversed in such a way that no node has more than one direct ancestor. In our case, for each connected component of the DD graph, we may start from any node as use either depth-first or breadth-first search to obtain the traversal ordering. During the traversal, each node is visited once. Fig.~\ref{fig:gateset_acyclic} depicts several idles and their traversal order. 

Below, we show that for a given traversal, each node with zero or one direct ancestor nodes can be always embedded optimally with exactly two X-gates (the minimum number of gates to resolve the identity operation). That is, we will show that two new X-gates are sufficient to refocus all phase and crosstalk errors to zero. This implies that acyclic DD embedding graphs can always be solved using the minimum two X-gates per idle, with all errors described by Eq.~\eqref{eq:hamiltonian} refocused exactly.

\subsection{DD Embedding in acyclic graphs}
\label{subsec:acyclic}

We embed DD by traversing a graph, embedding new X-gates into each node in turn. In this subsection, we show that if the current node in the traversal is connected to zero or one previously-visited nodes (direct ancestors in the graph traversal), then we can analytically determine the positions of two new X-gates in the current node to exactly suppress all error terms. 

First we consider a current node with zero immediate ancestors (such as the first node in the traversal).  In this case only the single-qubit phase is relevant as the lack of ancestor nodes indicates a lack of relevant crosstalk channels to be considered. This is easily suppressed using two X-gates spaced 50\% apart, for example at the start and halfway through the idle duration (see Fig.~\ref{fig:errors}a). Therefore, in this subsection, we focus on the one-ancestor case.

Consider the evolution of Eq.~\eqref{eq:hamiltonian} over the duration of a particular idle delay (called the `current' idle). Suppose that there is one further idle delay (the `ancestor' idle) that has a crosstalk interaction with the current idle. We wish for the qubit state at the end of the current idle to exactly match the state at the beginning of the current idle, so that the delay accurately realizes an error-free identity gate. The dynamics of the current idle before adding any DD are described by 
\begin{align}
U_{\text{idle}} = \exp(-i ( \epsilon \, t  \, IZ + \tau J ZZ))
\end{align}
where the qubit ordering of the Pauli operators is ancestor-current, $t$ is the current idle delay duration, $\epsilon$ and $J$ are the rates of phase and crosstalk accrual (as in Eq.~\eqref{eq:hamiltonian}). $\tau$ depends on the amount of time the ancestor is overlapped with the current idle, as well the positions of any X-gates already embedded on the the ancestor, which each reverse the sense of $ZZ$ accumulation and in general already produce some partial refocusing of crosstalk (if any are located within the mutually idle portion).  

Since all terms in Eq.~\eqref{eq:hamiltonian} commute, we can consider the evolution of $IZ$ and $ZZ$ separately. Two new gates separated by 50\% of the duration are sufficient to refocus $IZ$, as in Fig.~\ref{fig:errors}a. In fact, the offset of these gates (distance of the first gate from the start of the idle) can also be used to refocus $ZZ$, for any possible configuration of the overlap between the idles, and regardless of any DD gates already present on the ancestor. To see this, split the dynamics into two portions: up to the midpoint of the current idle ($U_1$), and beyond the midpoint ($U_2$):
\begin{align*}
    U_j = \exp(-i\tau_j J ZZ), ~~~ j \in \{1,2\}
\end{align*}
where $\tau = \tau_1 + \tau_2$. Note that generally $\tau_1 \neq \tau_2$. The $\tau_j$ may even have opposite sign, depending on the details of the arrangement of decoupling gates on the ancestor idle. We can represent an X-gate on the current qubit using the $IX$ operator. If we choose an offset of zero, then the two gates occur at the beginning and midpoint of the current idle, and the total evolution is:
\begin{align*}
    U &= U_2\,IX\,U_1\,IX = \exp(-i\Delta JZZ)
\end{align*}
where $\Delta  = \tau_2 - \tau_1$
Alternatively, if we choose offset of 50\%, then the two X-gates occur at the midpoint and endpoint of the current idle, and the total evolution is:
\begin{align*}
    V &= IX\,U_2\,IX\,U_1 = \exp(-i(-\Delta)JZZ) = U^{-1}
\end{align*}
Therefore, the total accumulation of crosstalk error between these two schemes differ by a sign change. However, we can also move smoothly between these two schemes by sliding the offset continuously from 0\% to 50\% of the idle duration. The value of total $ZZ$ phase responds continuously as we vary the offset, therefore there exists at least one zero-crossing. At that offset, the crosstalk is perfectly refocused and exactly suppressed  for any value of $J$. Note that the value of $\tau$ (and therefore of $\tau_1$ and $\tau_2$) depend on the precise number and arrangement of decoupling gates on the ancestor idle, as well as the timing of the mutually idle duration (which may be different from the full current idle duration). The important result here is that inserting two new X-gates on the current idle is always sufficient to realize an error-free identity operation, regardless of the intricacy of the configuration, and for any parameters of Eq.~\eqref{eq:hamiltonian}.

\begin{figure}
\centering
\includegraphics[width=0.95\linewidth]{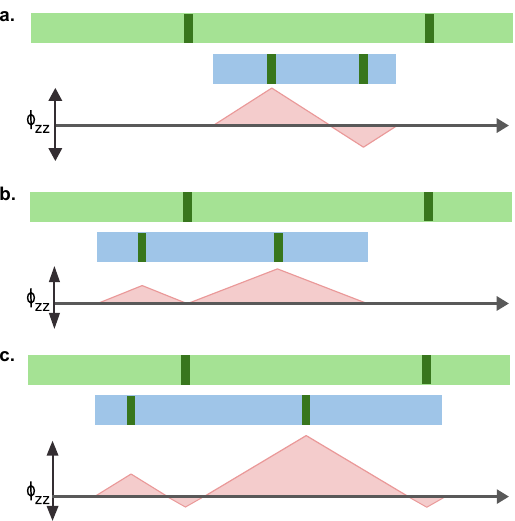}
\caption{\label{fig:theory_subinterval} Embedding of new gates into the current idle (blue) when it is a sub-interval of the constraint idle (green). The two X-gates added to the current idle should be separated by half the current idle duration in order to refocus phase error. The offset between the start of the current idle and the first gate can then be used to refocus crosstalk error. Red polygons depict the accumulation and refocusing of crosstalk phase during the mutually-idle period. (a) The constraint's gates are not inside the current idle duration. The two new gates may be embedded with any offset. (b) One constraint gate is inside the current idle. New gates should each be placed halfway between the current idle endpoint and the existing gate. (c) Both constraint gates are inside the current idle. One new gate should be placed halfway between the existing gates, the other should be 50\% of the current-idle duration away (either earlier or later). }
\end{figure}

For any particular configuration, one may analytically determine the desired offset based on the various idle start/stop times and ancestor X-gate locations. Specifically, the two gate times must be chosen such that the duration of forward error accumulation is exactly equal to the duration of reversed error accumulation. For example, Fig.~\ref{fig:theory_subinterval} contains the explicit algebraic conditions and solutions for all cases in which the current idle is a sub-interval of its direct ancestor (with the ancestor having two DD gate positions already determined). Similar expressions exist for other cases of interest; for example, when there is partial overlap between current and ancestor idles, or when the ancestor is a sub-interval of the current delay. 

For implementation purposes, we have explicitly solved all possible cases involving two X-gates on the direct ancestor, so that current idle offset can be immediately determined from its ancestor using simple algebraic conditions and expressions. The limitation to two gates is in a sense `closed'; since we always solve the current idle using two X-gates, we never need to solve a case involving a higher number of ancestor X-gates for a subsequent idle. Nonetheless, cases involving more than two ancestor DD gates can still be solved efficiently using any simple one-parameter root-finding algorithm. This is because the zero-crossing must exist, the number of new gates required is known to be two, and their relative spacing is fixed to 50\% of the current idle duration.

The argument above generalizes to cases of $n \geq 2$ equally spaced gates, each of which is $\pi$-pulse (not necessarily an X-gate). For example, a sequence comprising $n=4$ equally spaced gates can be continuously (and rigidly) translated between a configuration with a gate at the beginning, and another configuration with a gate at the end. These two arrangements generate opposite crosstalk phase, so there exists a zero-crossing for some offset partway between these two cases. Therefore, a large family of more intricate DD motifs enabling rigid translation of DD subsequences in time (such as XY4~\cite{Viola1999}, CPMG~\cite{meiboom1958}, KDD~\cite{souza2011robust}, Walsh DD~\cite{Hayes2011, Khodjasteh2013}, and so on) are also suitable as base embedding sequences used in our graph-traversal approach, in order to completely refocus crosstalk as well as phase errors.

\subsection{DD embedding graphs with cycles}
\label{subsec:cycles}

\begin{figure}
\centering
\includegraphics[width=0.95\linewidth]{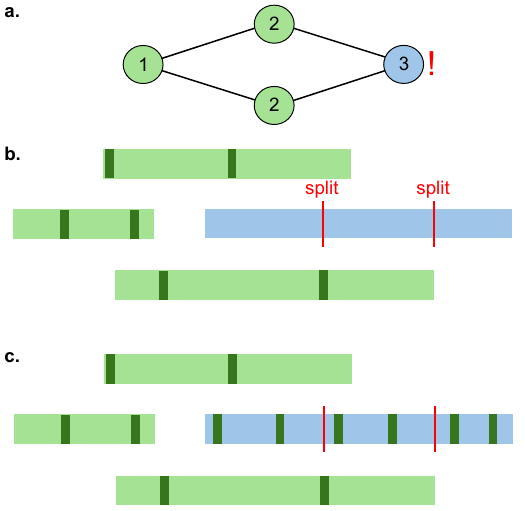}
\caption{\label{fig:cyclic} (a) A cycle in the DD embedding graph means that any traversal will eventually reach a node that has more than one direct ancestor. For example, the node labeled `3' depends on the timing and gate positions of {\em both} nodes labeled `2'. In the presence of more than one constraint, there is no analytical solution involving only two new X-gates. (b) An example of a node (blue) that is constrained by two already-embedded nodes (green). The multi-constrained node must be split into a series of shorter delays. (c) Each new node's DD embedding can be calculated based on at most one ancestor.}
\end{figure}

If the DD embedding graph contains cycles, then it is not possible to traverse the graph in such a way that each node has at most one direct ancestor. Instead, any traversal will inevitably visit nodes that have two or more direct ancestors, as depicted in Fig.~\ref{fig:cyclic}a. In general, there will be at most as many such multi-ancestor nodes as there are cycles in the graph.

Our method solves this problem by setting aside a set of nodes, such that the remaining subgraph is acyclic. A set of nodes that can be removed to leave an acyclic graph is called a feedback vertex set (FVS), and it is not unique. Determining the minimal FVS (i.e. the FVS with fewest nodes) is known to be an NP-complete problem, but heuristics can deliver suitable results.  We find an FVS simply by collecting nodes that have more than one immediate ancestor in the graph traversal. Since all the other nodes must have zero or one ancestors, the collected nodes indeed constitute an FVS. Other methods for finding an FVS could be used with the other parts of GraphDD. 
Regardless of the choice of FVS, each node with more than one direct ancestor must be eventually addressed. Our approach is to visit these nodes {\em last} so that all other nodes, including all their immediate neighbors, have already had their embedding determined.

FVS nodes are addressed by splitting the idle into several shorter idles, each receiving two X-gates, delivering an exact solution for the small penalty of an increase in decoupling gates from $2\to2n$, where $n$ (integer) is the number of sub-intervals that were introduced. Our approach to splitting is depicted in Fig.~\ref{fig:cyclic}b-c. Passing over the multi-constrained idle from left to right, split-points are introduced whenever the constraints go beyond what can be solved analytically. In this way, all sub-intervals (between the split points) are solved analytically using two new X-gates. The phase and crosstalk errors are refocused exactly over each sub-interval. In particular, by the time the rightmost edge of the idle is reached, all phase and crosstalk errors are suppressed exactly.

\subsection{Splitting of long idles at context changes}
\label{subsec:split_long_idles}

In circumstances with idles sufficiently long to violate the quasi-static assumption in Eq.~\eqref{eq:hamiltonian}, the rates of error accumulation may not be time-invariant over the idle duration.   Our approach to mitigating this circumstance is to split the long idle period into several smaller idles, each requiring a two-pulse DD sequence to be embedded.  The particular threshold for defining a ``long'' idle period is device-specific (or even qubit-specific) and may be set as a fraction of the average $T_2$ for the device.

Splitting the long idle into several smaller idles means that two X-gates will be embedded into each smaller idle. Therefore, the overall frequency of error refocusing is increased, compared to using two X-gates for the entire long idle. This approach is equivalent to the recognition that tuning the timing of interpulse periods can change the order of temporal noise variation cancelled by a DD sequence~\cite{Szwer_2011, Hayes2011}. For instance the repeated multipulse Carr-Purcell sequence is strictly able to cancel linear variation in the noise.   Further, such ``long-time'' sequence-construction considerations have been previously treated in general DD literature~\cite{Khodjasteh2013, Paz-Silva_2016}. 

\begin{figure}
\centering
\includegraphics[width=0.95\linewidth]{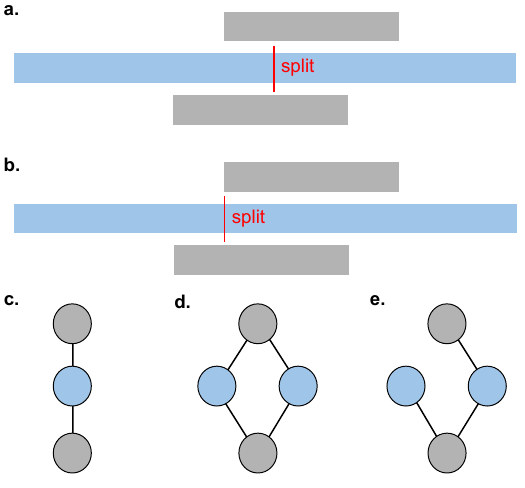}
\caption{\label{fig:split_long} Exceptionally long idle delays can be split into a series of shorter delays to increase the frequency of error refocusing. This increases the number of nodes in the DD embedding graph. (a) A long interval (blue) is split at an arbitrary point. (b) The interval is split at a context change, where a neighboring qubit becomes idle or ceases to be idle. (c) The original graph, prior to splitting. (d) An arbitrary split can generate a new cycle in the graph. (e) A split at a context change does not induce a new cycle. Since acyclic graphs are most easily solved, splitting at context changes is preferred.
}
\end{figure}

The key question is how the introduction of any form of long-idle splitting procedure is captured in the embedding procedure.  As shown in Fig.~\ref{fig:split_long}, the split can introduce new cycles into the DD embedding graph. However, additional cycles can be avoided if split points are chosen at `context changes'---moments at which other idles begin or end. Our implementation identifies an idle that is too long (according to a threshold) and calculates how many shorter idles it should be split into; if there are context changes sufficiently close to the ideal split points, then these are selected as the split points. However, where there are no suitable context changes, then the long idle is split at arbitrary points. This process introduces new context changes, where the new sub-intervals begin and end. Therefore, in cases of adjacent long intervals, one long idle may be split at arbitrary points, and the next will be split at the new context changes that were introduced. Therefore, the two (or more) adjacent long idles tend to be split at the same points in time.

\subsection{GraphDD algorithm overview}

The complete workflow of GraphDD is as follows:
\begin{enumerate}
    \item Accept a circuit: Given a compiled, scheduled quantum circuit and the quantum computer on which it will be executed, extract the timing and qubit connectivity information.

    \item Graph representation: Determine the timings of the idle delays on the various qubits. Based on the device connectivity, represent the DD embedding problem as a graph. See subection \ref{subsec:DD_as_a_graph}.

    \item Partition of long idles: Delays with duration greater than a predefined threshold are split into two or more smaller delays. See subsection \ref{subsec:split_long_idles}.
    
    \item Graph traversal: Use the structure of the graph to determine the order in which to add DD (i.e. additional gates or pulses) into the idles. Specifically, use a breadth-first search to generate a graph traversal.  See subsection \ref{subsec:DD_as_a_graph}.

    \item Embedding on nodes with zero or one direct ancestors: Determine the optimal gate locations for each node based on its direct ancestor node in the graph traversal. See subsection \ref{subsec:acyclic}.
    
    \item Splitting of nodes with two or more direct ancestors:  For the remaining set of nodes with more than one ancestor, determine the most favorable place(s) to split the idle duration into sub-intervals. These splits generate multiple new nodes and edges. Each new node has at most one direct ancestor. Determine the gate positions for each node based on its ancestor. See subsection \ref{subsec:cycles}.

    \item Circuit modification: Return a modified circuit that includes the embedded DD gates or pulses at the calculated times (subject to minor adjustments to obey device timing constraints). The other gates, and logical structure, are unchanged from the input circuit. 
\end{enumerate}

These steps are summarized as pseudocode in Algorithm~\ref{alg:graphDD} in order to highlight the relevant workflow and dependencies.

\begin{figure}
\begin{algorithm}[H]
\caption{GraphDD}\label{alg:graphDD}
\begin{algorithmic}[1]
    \State \textbf{Input}: scheduled circuit, connectivity map, max-idle time, device timing constraints
    \State $G \leftarrow$ Construct the DD graph
    \If {node in $G$ has duration $\geq$ max-idle}
        \State Split idle and update $G$
    \EndIf
    \State  $S\leftarrow$ graph traversal ordering according to BFS
    \If {node in $G$ has $\geq 2$ direct ancestors}
        \State Remove node from $G$
        \State FVS $\leftarrow$ save removed node 
    \EndIf
    \While {not all nodes in $G$ visited}
        \State $n \leftarrow$ pick a node according to $S$
        \State Add two X gates to $n$ based on $\text{ancestor}(n)$
    \EndWhile
    \For {node in FVS}
        \State Split node into several sub-intervals 
        \State $n\leftarrow$ Add new nodes to $G$
        \State Add two X gates to $n$ based on ancestor$(n)$
    \EndFor
    \State Adjust all gate times to obey device timing constraints
    \State Insert DD gates into the circuit at calculated times
    \State \textbf{Output}: scheduled circuit with embedded DD 
\end{algorithmic}
\end{algorithm}
\end{figure}

\section{Experimental benchmarking results}
\label{sec:performance}

\begin{figure*}
\centering
\includegraphics[width=0.95\linewidth]{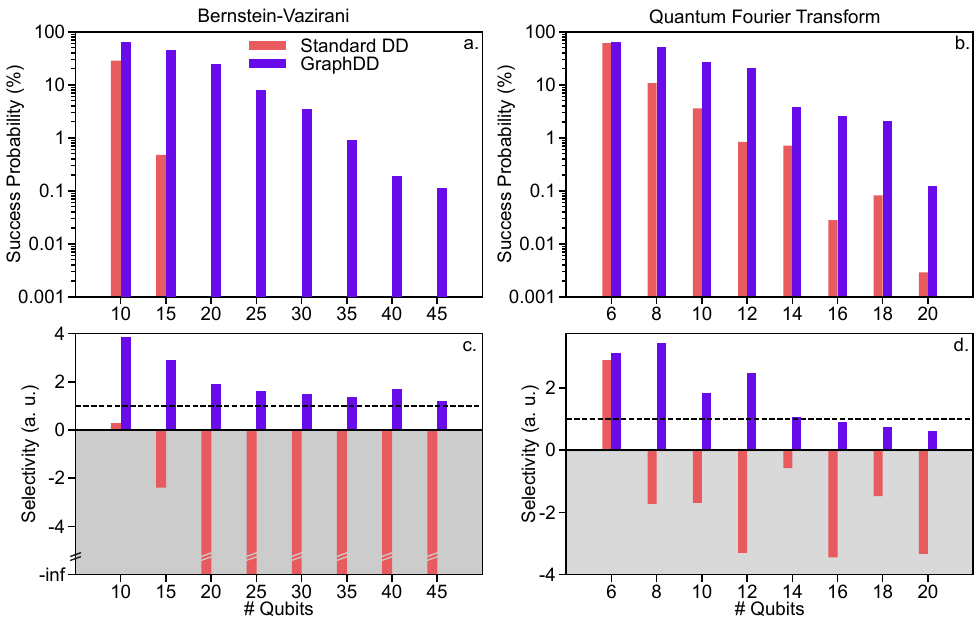}
\caption{\label{fig:result} Experimental comparison of GraphDD with a commonly used decoupling embedding scheme (standard). Data presented here were collected on the 127-qubit IBM device {\em ibm\_brisbane}. All executions have identical runtime settings.  Experiments employ 32,000 shots per circuit.
(a-b) Success probability (log scale) for the Bernstein-Vazirani (all `1' target) and quantum Fourier transform algorithms. The two circuits of each width differ {\em only} by their DD embedding.  (c-d) Selectivity of the correct result for the Bernstein-Vazirani algorithm and quantum Fourier transform algorithms.
}
\end{figure*}

\begin{figure*}
\centering
\includegraphics[width=0.95\linewidth]{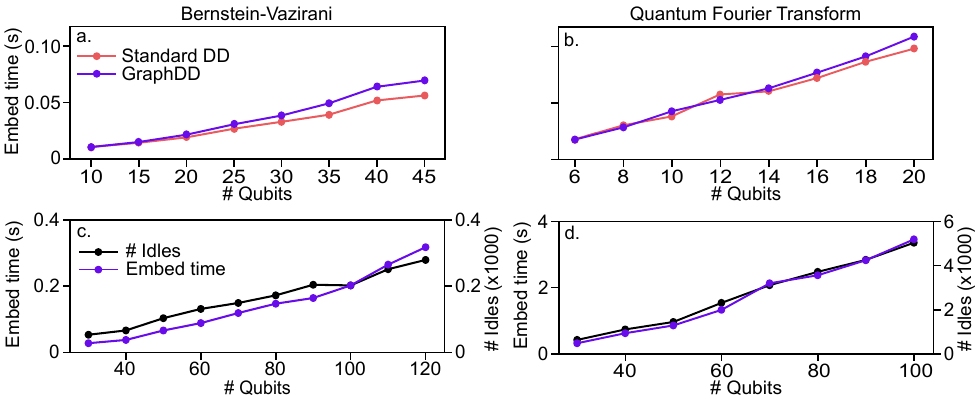}
\caption{\label{fig:compute} Classical computation time to embed DD into a circuit with GraphDD. (a-b) Compute time to embed DD in the same circuits that were used in Fig.~\ref{fig:result}, using GraphDD versus the standard embedding strategy. (c-d) Scaling of GraphDD embed time versus number of idles for large circuit widths, for BV and QFT, respectively.
}
\end{figure*}

We now demonstrate and quantify the impact of GraphDD in the execution of benchmark quantum circuits run on the 127-qubit processor {\em ibm\_brisbane}.  We analyze the performance of two widely-known quantum algorithms routinely used in benchmarking \cite{Mundada_2023,Wright_2019,hashim2024}, over a range of circuit widths up to the current state-of-the-art. 

In these circuit-level experimental tests we compare the performance of GraphDD with a ``Standard DD'' embedding which corresponds to the native Qiskit implementation of DD \cite{IBM_DD}. The standard sequence comprises two X-gates inserted into each delay, spaced at 25\% and 75\% of the idle duration, akin to the Carr-Purcell sequence. This refocuses phase errors exactly for static and linearly varying noise processes, but does so without consideration of circuit context. Accordingly, some crosstalk errors may be partially refocused by chance, but in general the crosstalk suppression is far from optimal. 

We ensure that the test circuits differ {\em only} by their DD embedding.  This is achieved by designing and transpiling the test circuit, and then making two identical copies, each receiving a different embedding of DD. All circuits are then submitted to the hardware at the same time (within the same job) and with identical runtime settings. Data are collected, processed, and presented using the same workflow. Identical measurement-error mitigation is applied to the raw counts for all circuits. Therefore, any observed difference in performance is due {\em solely} to the decoupling scheme in use.

For our benchmarks we select the Bernstein-Vazirani (BV) algorithm and Quantum Fourier Transform (QFT) because they possess different circuit characteristics and are likely to sample a variety of relevant contexts. For instance, BV uses relatively few entangling gates, but has long periods of mutual idling between neighboring qubits. The QFT is much more densely packed with two-qubit gates and more idle delays, yet the idles tend to have shorter duration.

BV finds a ``secret'' bitstring using only one query. We choose the ``all 1'' bitstring, since this requires the most entangling gates. The QFT performs a certain linear transformation of input states, and serves as a subroutine in many other quantum algorithms. The QFT can be made {\em one-hot}, i.e., an input layer of single qubit gates is used to ensure that the ideal output is a unique (arbitrary) bitstring. In this case, the target bitstring is an alternating pattern 1010...10.

For each algorithm we extract two measures of performance~\cite{Mundada_2023}. The first is the success probability of the algorithm; i.e., the frequency with which the correct result is obtained.  This provides an un-normalized measure of the likelihood that the correct output is returned, and naturally diminishes with increasing circuit width as the range of possible outputs grows. Secondly, we introduce the {\em selectivity} of the result. The selectivity is a measure of signal-to-noise, and it captures how much the correct result stands out above the other (incorrect) outputs. Selectivity is defined by:
\begin{align}
    s := \log_2\left(\frac{p_{\text{correct}}}{{p_{\text{next}}}}\right)
\end{align}
where $p_{\text{correct}}$ and $p_{\text{next}}$ are the observed likelihoods of the correct result and the most frequently obtained incorrect result, respectively. With positive selectivity, the most frequent result will eventually match the correct bitstring if the circuit is repeated enough times. Selectivity greater than one is defined as `strongly' selective, meaning the correct result is obtained substantially more often than any other bitstring. A negative selectivity means that the circuit is more likely to output an incorrect result; increased repetition or averaging over shots will not enable the correct result to emerge as the most-frequently observed bitstring.  

The results comparing both DD techniques across the two benchmark algorithms are presented in Fig.\ref{fig:result}. For the BV algorithm, we observe substantially better performance when using GraphDD, with at least $9,000\times$ higher success probability at 20 qubits (assuming that zero observed success likelihood corresponds to an underlying success probability of at most 1 / \# shots). Fig. \ref{fig:result}a and \ref{fig:result}c show that both success probability and selectivity remain high for widths of up to 45 qubits. For standard DD, in contrast, the circuit fails at 15 qubits and above; the correct bitstring is {\em not} the most frequently observed result ($s < 0$). In fact, for most circuit widths we test, the correct bistring is {\em never} observed amongst the 32,000 shots when using standard DD ($s = -\infty$). In contrast, the correct bitstring is the most frequently observed ($s>0$) in all cases when using GraphDD.

Fig.~\ref{fig:result}b and \ref{fig:result}d show the performance for the QFT algorithm. Maximum improvement in success probability is observed at $\sim 200\times$ for a 16-qubit implementation.  Beyond 6 qubits, only the GraphDD embedding reliably returns the correct answer as the most frequently observed result ($s>0$), and selectivity remains positive up to the maximum circuit width of 20 qubits tested. Standard DD does not produce the correct mode bitstring for most circuit sizes in this range ($s<0$).

We also record the classical compute time used to embed DD into each circuit, using the two methods, in order to provide insights into the computational overhead incurred in using GraphDD. Fig.~\ref{fig:compute}a-b show that the compute time required for GraphDD versus standard DD is approximately equal, for both the BV and QFT algorithms. At larger circuit widths (beyond the scale that produces meaningful experimental results on currently available devices), the scaling of classical compute time approximately follows the number of idle delays in the circuit (Fig.~\ref{fig:compute}c-d). 

We emphasize that these results are representative rather than specially selected.  We observe similar data on other devices (including other device sizes and coupling topologies), and for other algorithms and circuit widths.

\section{Discussion and Conclusion}

The method we present in this work, GraphDD, is a complete, automated solution to the DD embedding problem into arbitrary quantum circuits. It produces high-quality, circuit-optimized decoupling sequences for arbitrary input circuits and hardware topologies. Decoupling gate timing is determined using an analytic calculation (with no need for numeric optimization). It is efficient in required pulse resources, required pulse-timing resolution, and computational overhead. 

GraphDD completely refocuses quasi-static phase and crosstalk errors (up to minor device-timing constraints). For phase errors, the refocusing is exactly as effective as the simple default scheme that places gates at 25\% and 75\% of the way through each idle, according to full simulation of Eq.~\eqref{eq:hamiltonian}, accounting for all device timing constraints. However, GraphDD {\em also} achieves perfect suppression of crosstalk errors (up to device timing constraints) for arbitrary circuit structures. 

In arriving at GraphDD, we present several novel insights into the DD embedding problem.  First, an ordering of idles can be determined from the graph structure of the embedding problem. This dramatically simplifies calculation of optimal solutions. Second,  we find that efficient solutions exist for all possible configurations for how to embed DD on a node with zero or one direct ancestors in the graph traversal. In all possible configurations that could arise on any arbitrary input circuit, two single-qubit X-gates are sufficient to jointly refocus phase and crosstalk, regardless of the (unknown) parameters of Eq.~\eqref{eq:hamiltonian}, the timing of the idles, and the number and arrangement of DD gates on the ancestor idle.  Moreover, any $2n$ equally-spaced gates could be used if desired; for example, to accommodate a more intricate decoupling sequence such as XY4 or Walsh.  Third, handling of exceptionally long idles can be accommodated using splitting procedures in which the idle is partitioned at context-change points (where neighboring idles begin or end) and each sub-interval can be treated as a separate node in the embedding graph. And fourth, cycles in the embedding graph can be removed by analytically splitting one node based on the timings and gates of an arbitrary number of traversal direct ancestors, reducing the problem to an acyclic graph.

In GraphDD, the optimal embedding for each node can be determined analytically in all cases; there is never a need to numerically optimize the gate timings. This is achieved via a complete set of special cases that we have solved for each possible configuration of ancestor versus current node. This is an important practical consideration. Even for currently available devices, a very large number of idles can occur within a single circuit. And this number will increase as larger devices become available, and deeper circuits become feasible. Numerical optimization of DD embedding could easily form a bottleneck in the compilation process. This is especially true of joint optimization of parameters describing DD on many idles, which would form a high-dimensional optimization problem. In contrast, simple algebraic expressions that depend on the timing parameters of only two idles lead to an embedding process that scales linearly in the number of idle delays in the circuit. 

GraphDD is computationally efficient to implement, scaling approximately linearly with the number of idle delays in the input circuit. The main reason for the efficiency is that each node (idle delay) is only visited once, its gate positions are determined once, and they are not subsequently altered. In this way, a large number of independent parameters (the offsets for each idle delay in the circuit) are systematically optimized in series, rather than jointly optimized in a high-dimensional problem.  This is only possible because of the novel approach to ordering the idles; each idle must be addressed in turn based on its position in a graph traversal.  The linear scaling observed with the number of idles and the efficient embedding time per idle, supports the scalability of GraphDD well beyond near term applications. 

The utility of this technique is validated through benchmarking experiments on real quantum computing devices. In fact, the results achieved using GraphDD are, as far as we are aware, better than any other benchmark results on the same devices. The ability to automatically embed DD into arbitrary circuits and reliably obtain a substantial improvement in real-device fidelity is the primary advantage of this method.  

Finally, this method of embedding DD has been validated externally via third-party use in the error-suppression software pipeline Fire-Opal~\cite{Mundada_2023} since December 2023. GraphDD has contributed, along with other parts of that error-suppression pipeline, to a variety of impressive demonstrations of the capabilities of currently available quantum computing hardware in applications from optimization to quantum machine learning \cite{yamauchi2024,sachdeva2024,kanno2024}. We are excited by the fact that GraphDD's computational efficiency makes it scalable beyond current state-of-the-art devices, suggesting these early successes may be exceeded in future demonstrations accessing larger hardware systems.

\subsection*{Acknowledgments}
The authors are grateful to all other colleagues at Q-CTRL whose technical, product engineering, and design work has supported the results presented in this paper.

\bibliography{refs}

\end{document}